\documentclass[aps,prd,superscriptaddress,showpacs,showkeys,floatfix,twocolumn,nofootinbib]{revtex4}

\usepackage{graphicx}
\usepackage{dcolumn}
\usepackage{amsmath,amssymb,bm}
\usepackage{citesort}

\newcommand{\be}{\begin{equation}}
\newcommand{\ee}{\end{equation}}
\newcommand{\epTc}{\eta'\to 2(\pi^+\pi^-)}
\newcommand{\epTcn}{\eta'\to \pi^+\pi^-2\pi^0}
\newcommand{\epTn}{\eta'\to 4\pi^0}
\newcommand{\eTn}{\eta\to 4\pi^0}
\newcommand{\mep}{M_{\eta'}}
\newcommand{\Amp}{\mathcal{A}}
\newcommand{\F}{\mathcal{F}}
\newcommand{\G}{\mathcal{G}}
\newcommand{\BR}{\mathcal{B}}
\newcommand{\Order}{\mathcal{O}}
\newcommand{\Lagr}{\mathcal{L}}
\newcommand{\nnnl}{\nonumber\\}
\newcommand{\half}{\textstyle\frac{1}{2}}
\newcommand{\eps}{\epsilon}
\renewcommand{\Im}{\textrm{Im}\,}

\allowdisplaybreaks[1]

\begin{document}
%--------------------------------------------------------------------------------

\title{Anomalous decays of \boldmath{$\eta'$} and \boldmath{$\eta$} into four pions}

\author{Feng-Kun Guo}
\email{fkguo@hiskp.uni-bonn.de}
\affiliation{Helmholtz-Institut f\"ur Strahlen- und Kernphysik (Theorie) and\\
             Bethe Center for Theoretical Physics,
             Universit\"at Bonn,
             D-53115 Bonn, Germany}

\author{Bastian Kubis}
\email{kubis@hiskp.uni-bonn.de}
\affiliation{Helmholtz-Institut f\"ur Strahlen- und Kernphysik (Theorie) and\\
             Bethe Center for Theoretical Physics,
             Universit\"at Bonn,
             D-53115 Bonn, Germany}

\author{Andreas Wirzba}
\email{a.wirzba@fz-juelich.de}
\affiliation{Institute for Advanced Simulation
         and J\"ulich Center for Hadron Physics,
             Institut f\"ur Kernphysik,
             Forschungszentrum J\"ulich,
             D-52425 J\"ulich, Germany}

\date{\today}

\begin{abstract}
We calculate the branching ratios of the yet unmeasured $\eta'$ decays
into four pions, based on a combination of chiral perturbation theory and
vector-meson dominance. The decays $\epTc$ and $\epTcn$ are P-wave dominated and
can largely be thought to proceed via two $\rho$ resonances; 
we predict branching fractions of $(1.0\pm 0.3) \times 10^{-4}$ and $(2.4 \pm
0.7) \times 10^{-4}$, respectively, not much lower than the current experimental upper limits.
The decays $\epTn$ and $\eTn$, in contrast, are D-wave driven as long as
conservation of $CP$ symmetry is assumed, and are significantly further suppressed; 
any experimental evidence for the decay $\eTn$ could almost certainly be interpreted as
a signal of $CP$ violation.
We also calculate the $CP$-violating amplitudes for $\epTn$ and $\eTn$ induced by the QCD
$\theta$-term.
\end{abstract}



\maketitle

%--------------------------------------------------------------------------------
\section{Introduction}
%--------------------------------------------------------------------------------

Processes in low-energy QCD that involve an odd number of (pseudo-)Goldstone bosons
(and possibly photons), which are, therefore, of odd intrinsic parity, are thought
to be governed by the Wess--Zumino--Witten (WZW) term~\cite{WZW} via chiral anomalies.  
While the so-called triangle anomaly is well tested in processes such as 
$\pi^0,\,\eta\to\gamma\gamma$, 
and the box anomaly contributes e.g.\ to $\gamma\pi\to\pi\pi$ and $\eta\to\pi\pi\gamma$,
the pentagon anomaly remains more elusive; the simplest possible process that is usually
cited is $K^+K^-\to\pi^+\pi^-\pi^0$, which however has not been experimentally tested yet,
and is likely to be subject to large corrections to the chiral-limit amplitude that is
dictated by the WZW term.

A different set of processes involving five light pseudoscalars is the four-pion decays
of $\eta$ and $\eta'$.  Experimental information about these is scarce: only upper limits
on branching ratios exist~\cite{PDG2010}; however, this may change in the near future for at least
some of the possible final states with the advent of high-statistics $\eta'$ experiments
such as BES-III, WASA-at-COSY, ELSA, CB-at-MAMI-C, CLAS at Jefferson Lab, etc.
We are only aware of one previous theoretical calculation of these decays, 
performed in the framework of a quark model~\cite{Parashar},
whose partial width predictions, however, have in the meantime been ruled out by the experimental
upper limits, at least for the channel $\epTc$.

In principle, the decays $\eta' \to 4\pi$, in contradistinction
to many other $\eta'$ decay channels, seem not terribly forbidden by approximate symmetries:
they are neither isospin forbidden, nor required to proceed via electromagnetic interactions.
The reaction $\eta\to 4\pi$, in contrast, is essentially suppressed by tiny phase space: only the decay
into $4\pi^0$ is kinematically allowed ($M_\eta-4M_{\pi^0} = 7.9$\,MeV,
$M_\eta-2(M_{\pi^\pm}+M_{\pi^0}) = -1.2$\,MeV).
Furthermore, the fact that anomalous amplitudes always involve
the totally antisymmetric tensor $\eps_{\mu\nu\alpha\beta}$ can be used to show that no two pseudoscalars
are allowed to be in a relative S~wave:
assuming they were, this would reduce the five-point function $PPPPP$ effectively to
a four-point function $SPPP$ (where $S$ stands for a scalar and $P$ for a pseudoscalar), 
in which there are no four independent vectors left to contract
the $\eps$ tensor with.
The decays $\epTc$ and $\epTcn$ can therefore be expected to be P-wave dominated.
As furthermore Bose symmetry forbids two neutral pions to be in an odd partial wave,
$\epTn$ and $\eTn$ even require all $\pi^0$ to be at least in relative D~waves~\cite{KupscWirzba}.
This, combined with the tiny phase space available, leads to the notion of $\eTn$ being
$CP$ forbidden~\cite{PDG2010,Prakhov,Nefkens}, although strictly speaking it is only S-wave $CP$ forbidden.

The outline of the article is as follows. We begin by discussing the two decay
channels with charged pions in the final state, $\epTc$ and $\epTcn$, in
Sec.~\ref{sec:Charged}. There, we calculate the corresponding decay amplitudes at
leading nonvanishing order in the chiral expansion, saturate the  appearing low-energy
constants by vector-meson contributions, and calculate the
corresponding branching ratios. In Sec.~\ref{sec:Neutral}, we then construct a
$CP$-conserving (D-wave) decay mechanism for $\eta,\,\eta'\to 4\pi^0$ and determine
the resulting branching fractions, before discussing the $CP$-violating (S-wave)
$\eta,\,\eta' \to 4\pi^0$ decay as induced by the QCD $\theta$-term in Sec.~\ref{sec:CP}.
Finally, we summarize and conclude. The Appendices contain technical details on
four-particle phase space integration as well as on a (suppressed) tensor-meson
mechanism for $\eta,\,\eta' \to 4\pi^0$.

%--------------------------------------------------------------------------------
\section{\boldmath{$\epTc$ and $\epTcn$}}
\label{sec:Charged}
%--------------------------------------------------------------------------------
\subsection{Chiral perturbation theory}
%--------------------------------------------------------------------------------

We wish to calculate the leading (nontrivial) chiral contribution to the anomalous decays
\begin{align}
\eta' &\to \pi^+(p_1)\pi^-(p_2)\pi^+(p_3)\pi^-(p_4) ~, \nnnl
\eta' &\to \pi^+(p_1)\pi^0(p_2)\,\pi^-(p_3)\pi^0(p_4) ~. \label{eq:charged}
\end{align}
The amplitudes can be written in terms of the invariant variables $s_{ij} = (p_i+p_j)^2$,
$i,\,j = 1,\ldots,4$,
which are subject to the constraint
\be
s_{12}+s_{13} +s_{14}+s_{23}+s_{24}+s_{34} = \mep^2 + 8M_\pi^2
\ee
(in the isospin limit of equal pion masses).
The five-meson vertices of the WZW term can be deduced from the Lagrangian
\be
\Lagr_{P^5}^{\rm WZW} =
\frac{N_c\eps_{\mu\nu\alpha\beta}}{240\pi^2F_\pi^5}
 \left\langle \varphi \partial^\mu\varphi
\partial^\nu\varphi \partial^\alpha\varphi \partial^\beta\varphi \right\rangle + \ldots,
\label{eq:WZWpentagon}
\ee
where $N_c$ is the number of colors and will be taken to be 3 in this paper, $F_\pi=92.2\,{\rm MeV}$
is the pion decay constant, and $\langle \ldots \rangle$ denotes the trace in
flavor space. For simplicity, we refrain from spelling out the WZW term in its
full, chirally invariant form. Furthermore, 
\be 
\frac{\varphi}{\sqrt{2}} = \left(
\begin{array}{ccc}
\frac{\eta_0}{\sqrt{3}} + \frac{\eta_8}{\sqrt{6}} + \frac{\pi^0}{\sqrt{2}} & \pi^+ & K^+\\[2mm]
\pi^- & \frac{\eta_0}{\sqrt{3}} + \frac{\eta_8}{\sqrt{6}} -\frac{\pi^0}{\sqrt{2}} & K^0\\[2mm]
K^- & \bar{K}^0 & \frac{\eta_0}{\sqrt{3}} -\frac{2\eta_8}{\sqrt{6}}
\end{array}
\right).
\ee
We assume a simple, one-angle $\eta\eta'$ mixing scheme,
\begin{align}
|\eta\rangle &= \cos\theta_P|\eta_8\rangle -\sin\theta_P |\eta_0\rangle ~, \nnnl
|\eta'\rangle &= \sin\theta_P|\eta_8\rangle +\cos\theta_P |\eta_0\rangle ~,
\label{eq:mixing}
\end{align}
and use the standard mixing angle $\theta_P = \arcsin({-1}/{3})\approx
-19.5^\circ$. As we are going to present what in some sense corresponds to a
leading-order calculation of the decay amplitudes, we regard the more elaborate
two-angle mixing schemes~\cite{LeutwylerKaiser} as beyond the scope of this study;
we expect the error made thereby to be covered by our generous final uncertainty
estimate.

\begin{figure}
\centering
\includegraphics[width=0.9\linewidth]{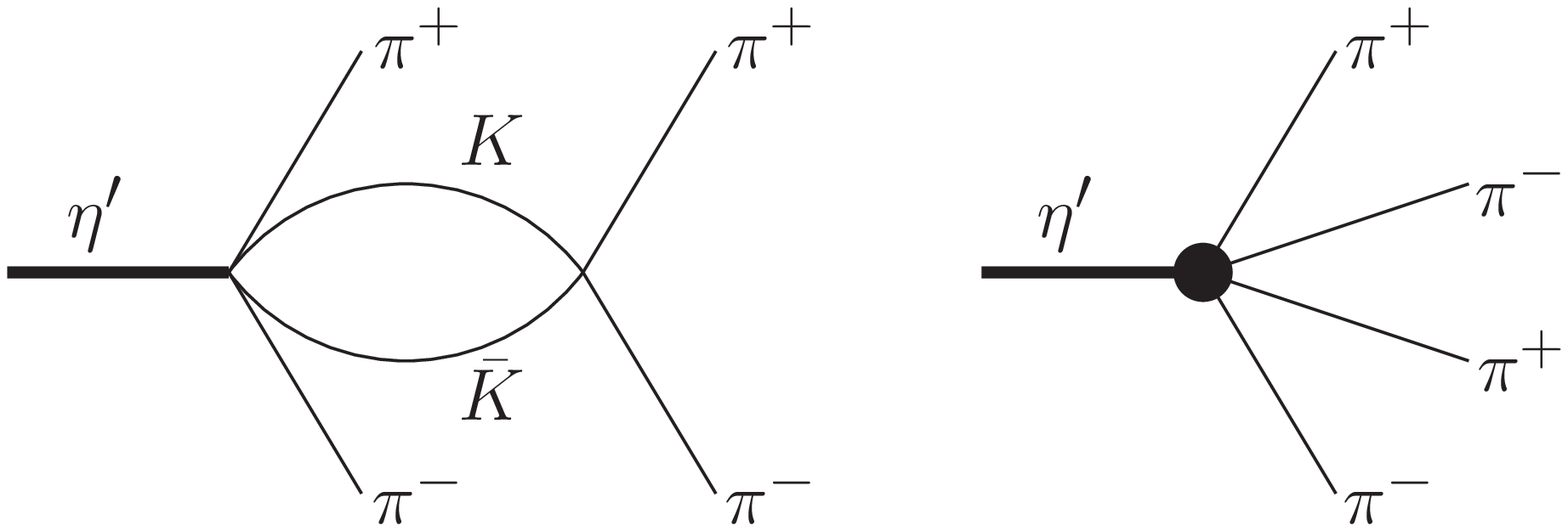}
\caption{Feynman diagrams contributing to $\epTc$ (and similarly to $\epTcn$) at $\Order(p^6)$.
The thick dot in the right diagram denotes a vertex from $\Lagr_{\rm odd}^{(6)}$.}
\label{fig:KKloop}
\end{figure}

The flavor structure of Eq.~\eqref{eq:WZWpentagon}
is such that there are no direct contributions to $\eta,\,\eta'\to 4\pi$,
and the decay amplitudes vanish
at leading order (in the anomalous sector) $\Order(p^4)$.
Nonvanishing contributions only
occur at $\Order(p^6)$, where the amplitudes are given by  sums
of (kaon) loops and counterterm contributions from the $\Order(p^6)$ Lagrangian of odd
intrinsic parity~\cite{BGT_Op6}, see Fig.~\ref{fig:KKloop}.
Only two different structures ($\propto C_1^W, \,  C_{12}^W$) remain when external currents are switched off.
Ref.~\cite{BGT_Op6} only considers the Goldstone boson octet;
we add terms $\propto \tilde C_1^W, \, \tilde C_{12}^W$ that only contribute for
the singlet field $\eta_0$:
\begin{align}
\Lagr^{(6)}_{\rm odd} &=
i C_1^W \eps_{\mu\nu\alpha\beta} \langle \chi_- u^\mu u^\nu u^\alpha u^\beta \rangle \nnnl
& - \frac{i \tilde C_1^W}{3} \eps_{\mu\nu\alpha\beta} \langle \chi_-\rangle\langle u^\mu u^\nu u^\alpha u^\beta \rangle
\nnnl
&+ C_{12}^W \eps_{\mu\nu\alpha\beta} \langle h^{\gamma\mu}
[u_\gamma, u^\nu  u^\alpha  u^\beta]\rangle \nnnl
& - \frac{\tilde C_{12}^W}{3} \eps_{\mu\nu\alpha\beta}
\langle h^{\gamma\mu} [u_\gamma, u^\nu  u^\alpha] \rangle \langle u^\beta \rangle
+ \ldots ~, \label{eq:LagrC12}
\end{align}
with the usual chiral vielbein $u_\mu = i(u^\dagger \partial_\mu u - u \partial_\mu
u^\dagger)$ (neglecting external currents), $u = \exp (i \varphi/2F_\pi)$,
$h_{\mu\nu} = \nabla_\mu u_\nu + \nabla_\nu u_\mu$ with $\nabla_\mu X = \partial_\mu X + [\Gamma_\mu,X]$ and
$\Gamma_\mu= \half ( u^\dagger \partial_\mu u+ u \partial_\mu u^\dagger)$ (neglecting again external currents). Furthermore,
we use  $\chi_- = u^\dagger\chi
u^\dagger - u \chi^\dagger u$, where $\chi = 2 B \,{\rm diag}(m_u,m_d,m_s) +
\ldots$ contains the quark mass matrix and $B$ is related to the quark
condensate according to $B=-\langle \bar q q\rangle/F_\pi^2$.
The decay amplitudes at $\Order(p^6)$ take the compact forms
\begin{align}
\Amp(\eta_{0/8}&\to\pi^+\pi^-\pi^+\pi^-) = - \Amp(\eta_{0/8}\to\pi^+\pi^0\pi^-\pi^0) \nnnl
&= \frac{N_c\eps_{\mu\nu\alpha\beta}}{3\sqrt{3}F_\pi^5} p_1^\mu p_2^\nu p_3^\alpha p_4^\beta
\big[ \F_{0/8}(s_{12})+\F_{0/8}(s_{34}) \nnnl
&\quad -\F_{0/8}(s_{14})-\F_{0/8}(s_{23}) \big] ~, \nnnl
\F_0(s) &= - 16 \sqrt{2}\left(C_{12}^{Wr}(\mu)-\tilde C_{12}^{Wr}(\mu)\right)s ~, \nnnl
\F_8(s) &= \frac{1}{8\pi^2F_\pi^2}\bigg\{ \big(s-4M_K^2\big) \bar J_{KK}(s) \nnnl
& \quad - \frac{s}{16\pi^2}
\Big( 2\log\frac{M_K}{\mu} + \frac{1}{3} \Big) \bigg\}
- 16 C_{12}^{Wr}(\mu)s ~, \nnnl
\bar{J}_{KK}(s) &=\frac{1}{8\pi^2}(1-\sigma_K\operatorname{arccot}\sigma_K)\,, ~
\sigma_K=\sqrt{\frac{4M_K^2}{s}-1} \,, \label{eq:AmpOp6}
\end{align}
with the scale-dependent renormalized low-energy constants $C_{12}^{Wr}(\mu)$ and $\tilde
C_{12}^{Wr}(\mu)$. There are no loop
contributions to the $\eta_0$ amplitudes at this order, since at
$\Order(p^4)$ the anomalous five-pseudoscalar term (\ref{eq:WZWpentagon}) 
 (the left vertex  of the loop diagram in Fig.~\ref{fig:KKloop}) contributes only to the octet case.
 Equation~\eqref{eq:AmpOp6}
is scale-independent with the $\beta$ function for $C_{12}^{Wr}(\mu)$ obtained in
Ref.~\cite{BGT_Op6}, if we demand $\tilde C_{12}^{W}$ to have the same infinite
part and scale dependence as $C_{12}^{W}$. A numerical estimate for the finite part
$C_{12}^{Wr}(M_\rho)$ will be obtained by resonance saturation through vector-meson
contributions.

%--------------------------------------------------------------------------------
\subsection{Resonance saturation from hidden local symmetry}
%--------------------------------------------------------------------------------

Resonance saturation for the $\Order(p^6)$ chiral Lagrangian of odd intrinsic parity
has been studied in great generality recently in Ref.~\cite{Kampf}.
Here however we opt for the simpler, but on the other hand more predictive
hidden-local-symmetry scheme~\cite{FKTUY,Ulf,Bando,Harada}, which has the additional advantage of having been
tested phenomenologically in great detail~\cite{Benayoun}.

In the framework of hidden local symmetry (HLS), there are four additional terms
involving vector-meson fields, with coefficients $c_i$ $(i=1,\ldots,4)$, in
addition to the WZW action for anomalous processes~\cite{FKTUY,Bando}; as already
noted in Ref.~\cite{BijnensBramonCornet}, only three independent combinations of
these contribute to low-energy amplitudes at $\Order(p^6)$. HLS amplitudes for any
given anomalous process contain two kinds of contributions: contact terms and
resonance exchange terms. The contact terms have the same form as those derived
from the WZW action, but with a modified coefficient (see below); the
gauge-invariant construction of the HLS Lagrangian density guarantees that the
additional, $c_i$-dependent contributions will be canceled by vector-meson exchange
in the low-energy limit. In the following, we again for simplicity reasons refrain
from properly defining all the HLS Lagrangian terms in their chirally
invariant forms, but only quote the terms relevant for vertices of five
pseudoscalars; the full Lagrangians can be retrieved e.g.\ from
Refs.~\cite{Bando,Harada}.

The contact terms for five-pseudoscalar vertices can be read off from the Lagrangian
\be
\Lagr_{P^5}^{\rm HLS} =
\frac{N_c\eps_{\mu\nu\alpha\beta}}{240\pi^2F_\pi^5}\left[1-\frac{15}{8}(c_1\!-\!c_2)\right]
 \left\langle \varphi \partial^\mu\varphi
\partial^\nu\varphi \partial^\alpha\varphi \partial^\beta\varphi \right\rangle .
\label{eq:HLScontact}
\ee
The low-energy limit of the
vector-meson-exchange contribution can be obtained by integrating out the heavy
fields: substituting the leading-order equation of motion of the vector-meson fields
\be
V_\mu = \frac1{8igF_\pi^2}\left[\partial_\mu\varphi,\varphi\right]  , \label{eq:EOMvector}
\ee
where $g$ is the universal vector-meson coupling constant, 
into the HLS Lagrangians~\cite{Harada,Benayoun}
\begin{align}
\Lagr_{VVP} &=  - \frac{N_c c_3 g^2}{8\pi^2F_\pi}\eps_{\mu\nu\alpha\beta}
\left\langle \partial^\mu V^\nu \partial^\alpha V^\beta \varphi \right\rangle ,\nnnl
\Lagr_{VPPP} &= - \frac{iN_c (c_1-c_2-c_3) g}{32\pi^2F_\pi^3}\eps_{\mu\nu\alpha\beta}
\left\langle V^\mu \partial^\nu\varphi\partial^\alpha\varphi\partial^\beta\varphi \right\rangle ,
\label{eq:HLSLagr}
\end{align}
where the vector-meson nonet (with ideal mixing) is defined as
\begin{align}
V_\mu = \frac{1}{\sqrt{2}}\left(
\begin{array}{ccc}
\frac{\rho^0_\mu}{\sqrt{2}} + \frac{\omega_\mu}{\sqrt{2}} & \rho^+_\mu & K^{*+}_\mu\\[2mm]
\rho^- & -\frac{\rho^0_\mu}{\sqrt{2}} + \frac{\omega_\mu}{\sqrt{2}} & K^{*0}_\mu\\[2mm]
K^{*-}_\mu & \bar{K}^{*0}_\mu & \phi_\mu
\end{array}
\right) ,
\end{align}
we find
\be
\Lagr_{P^5,V}^{(4)} = \frac{N_c(c_1-c_2)}{128\pi^2F_\pi^5}
\eps_{\mu\nu\alpha\beta} \left\langle \varphi \partial^\mu\varphi
\partial^\nu\varphi \partial^\alpha\varphi \partial^\beta\varphi \right\rangle,
\label{eq:HLSvec}
\ee
which exactly cancels the second term inside the square brackets in
Eq.~\eqref{eq:HLScontact}.

If we extend the equation of motion Eq.~\eqref{eq:EOMvector} to next-to-leading
order in the derivative expansion,
\be
V_\mu =
\frac1{8igF_\pi^2}\left(1-\frac{\partial^2}{M_V^2}\right)\left[\partial_\mu\varphi,\varphi\right]
~, \label{eq:EOMvectorNLO}
\ee
where $M_V$ is the vector-meson mass, we can derive the vector-meson contribution
to the five-meson vertices at $\Order(p^6)$. Inserting Eq.~\eqref{eq:EOMvectorNLO}
into Eq.~\eqref{eq:HLSLagr}, we find
\begin{align}
\Lagr_{P^5,V}^{(6)} &= \frac{N_c(c_1-c_2+c_3)}{128\pi^2F_\pi^5M_V^2} \eps_{\mu\nu\alpha\beta} \nnnl
& \times \big\langle \partial^\lambda\partial^\mu\varphi \left[\partial_\lambda\varphi,
\partial^\nu\varphi \partial^\alpha\varphi \partial^\beta\varphi\right] \nnnl
& \quad -2 \partial^2\varphi \partial^\mu\varphi \partial^\nu\varphi \partial^\alpha\varphi \partial^\beta\varphi
 \big\rangle ~.
\end{align}
The first term is exactly of the form of the Lagrangian term $\propto C_{12}^W$ in Eq.~\eqref{eq:LagrC12}.
For the second term, we use the equation of motion for the Goldstone bosons, which,
neglecting higher orders in the fields, reads (compare e.g.\ Ref.~\cite{BGT_Op6})
\be
\partial^2 \varphi = - \tfrac{1}{2}\{\varphi,\chi\} + \tfrac{1}{3}\langle\varphi\chi\rangle ~,
\ee
so we also identify a vector-meson contribution to $C_1^{Wr}$ and $\tilde C_1^{Wr}$.  Our results read altogether
\begin{align}
C_1^{Wr}(M_V) &= \tilde C_1^{Wr}(M_V) = -2 C_{12}^{Wr}(M_V) \nnnl
& = \frac{N_c(c_1-c_2+c_3)}{128\pi^2M_V^2} ~, \label{eq:C1+12ressat}
\end{align}
where we have indicated the conventional assumption of the resonance-saturation
hypothesis to be valid roughly at the resonance scale, $\mu=M_V$ (which in the
following we will identify with the mass of the $\rho$, $M_\rho = 775.5$\,MeV). The
numerical values of the HLS coupling constants are often taken to be given by
$c_1-c_2 \approx c_3 \approx1$~\cite{FKTUY}, fairly consistent with more elaborate
phenomenological fits that yield $c_1-c_2=1.21$,
$c_3=0.93$~\cite{Benayoun}.

In principle, this completes the task to provide the necessary input for an evaluation of
the chiral representation of the decay amplitude, Eq.~\eqref{eq:AmpOp6}.
We observe, however, the following.
First, evaluating the slope of (the largely linear function) $\F_8(s)$
in Eq.~\eqref{eq:AmpOp6} with this input (using $\bar J'_{KK}(0) =
1/(96\pi^2M_K^2)$), we find
\begin{align}
 & 8\pi^2 (4\pi F_\pi)^2 \times \F'_8(0) \nnnl
&= 3 (c_1-c_2+c_3) \frac{(4\pi F_\pi)^2}{2M_\rho^2}
- \Big(1+2\log\frac{M_K}{M_\rho}\Big) ~.  
\label{eq:VMvsKloop}
\end{align}
Numerically, the first term is about $6.7\times  (c_1-c_2+c_3)/2$, and the second is 0.1. Hence, at the
scale $\mu=M_\rho$, the slope is entirely dominated by the vector-meson
contribution, and the kaon loops are negligible.

Second, the maximal value for the kinematical invariants in $\eta' \to 4\pi$
allowed by phase space is $\sqrt{s_{ij}} \leq \mep-2M_\pi \approx 680$\,MeV,
therefore replacing the $\rho$ propagator by its leading linear approximation is
not phenomenologically reliable. Even deviations induced by the finite \emph{width} of the $\rho$,
$\Gamma_\rho = 149.1$\,MeV, will be clearly visible.  In the following, we will
therefore use the full vector-meson-exchange amplitudes as derived from the HLS
formalism, with the $\rho$-meson propagators including the width, which in addition
is expected to be a very good estimate of the higher-order pairwise P-wave
interaction of the pions in the final state (of course neglecting any
crossed-channel effects). They are given by
\begin{align}
&\Amp_{V}(\eta_8\to\pi^+\pi^-\pi^+\pi^-) = \frac{1}{\sqrt{2}} \Amp_{V}(\eta_0\to\pi^+\pi^-\pi^+\pi^-) \nnnl
&= - \Amp_{V}(\eta_8\!\to\pi^+\pi^0\pi^-\pi^0) = - \frac{1}{\sqrt{2}} \Amp_{V}(\eta_0\to\pi^+\pi^0\pi^-\pi^0) \nnnl
&= \frac{N_c\eps_{\mu\nu\alpha\beta}}{16\sqrt{3}\pi^2F_\pi^5}p_1^\mu p_2^\nu p_3^\alpha p_4^\beta
\bigg\{ (c_1-c_2-c_3)\bigg[\frac{M_\rho^2}{D_\rho(s_{12})} \nnnl
& \qquad + \frac{M_\rho^2}{D_\rho(s_{34})}
                     - \frac{M_\rho^2}{D_\rho(s_{14})} - \frac{M_\rho^2}{D_\rho(s_{23})} \bigg] \nnnl
& + 2c_3 \bigg[ \frac{M_\rho^4}{D_\rho(s_{12})D_\rho(s_{34})}
             - \frac{M_\rho^4}{D_\rho(s_{14})D_\rho(s_{23})} \bigg] \bigg\} \label{eq:Arho1}\\
&\simeq \frac{N_c\eps_{\mu\nu\alpha\beta}}{16\sqrt{3}\pi^2F_\pi^5}p_1^\mu p_2^\nu p_3^\alpha p_4^\beta
\bigg\{ (c_1-c_2)\bigg[\frac{s_{12}}{D_\rho(s_{12})} \nnnl  & \qquad + \frac{s_{34}}{D_\rho(s_{34})}
                     - \frac{s_{14}}{D_\rho(s_{14})} - \frac{s_{23}}{D_\rho(s_{23})} \bigg] \nnnl
&  + c_3 \bigg[ \frac{M_\rho^2(s_{12}+s_{34})}{D_\rho(s_{12})D_\rho(s_{34})}
             - \frac{M_\rho^2(s_{14}+s_{23})}{D_\rho(s_{14})D_\rho(s_{23})} \bigg] \bigg\}
~,\label{eq:Arho2}
\end{align}
where
\begin{align}
D_\rho(s) &= M_\rho^2 -s - i \,M_\rho\Gamma_\rho(s)  ~, \nnnl
\Gamma_\rho(s) &= \frac{M_\rho}{\sqrt{s}}\bigg(\!\frac{s-4M_\pi^2}{M_\rho^2-4M_\pi^2}\!\bigg)^{3/2}\Gamma_\rho
\end{align}
 is the inverse $\rho$ propagator, and
we have neglected the width term in the transformation from Eq.~\eqref{eq:Arho1}
to Eq.~\eqref{eq:Arho2}
in order to demonstrate the correct chiral dimension $\Order(p^6)$ of the vector-meson contribution explicitly.
Expanding the resonance propagators in Eq.~\eqref{eq:Arho2} and comparing to Eq.~\eqref{eq:AmpOp6}
easily leads back to the coupling constant estimate for $C_{12}^{Wr}$
found on the Lagrangian level in Eq.~\eqref{eq:C1+12ressat}.

At this point, 
we can try to answer the introductory question on which
parts of the WZW anomaly action---triangle, box, or pentagon---the decays
$\epTc$ and $\epTcn$ yield information.  As the pentagon anomaly only enters via
the kaon-loop contributions, we have found above that its significance for the decays
under investigation here is negligible; the vector-meson contributions are derived from 
the triangle and box-anomaly terms, see Eq.~\eqref{eq:HLSLagr}.
As the phenomenological values of the HLS coupling constants
suggest $c_1-c_2-c_3 \ll 2c_3$, the box anomaly yields the lesser part of the two, 
and the decays are dominated by the triangle-anomaly term.

%--------------------------------------------------------------------------------
\subsection{Branching ratios}
\label{sec:numCharged}
%--------------------------------------------------------------------------------

We calculate the partial widths of the decays $\eta'\to \pi^+\pi^-\pi^+\pi^-$ and
$\eta'\to \pi^+\pi^0\pi^-\pi^0$ using
\be
\label{eq:width} \Gamma(\eta' \to 4\pi) = \frac{1}{2S\mep} \int |\Amp(\eta'\to
4\pi)|^2 d\Phi_4 ~,
\ee
where the evaluation of the
four-particle phase space $\Phi_4$ is discussed in detail in Appendix~\ref{app:ps}.
$S$ is a symmetry factor---$S=4$ for the $2(\pi^+\pi^-)$ final state, and $S=2$
for the $\pi^+\pi^-2\pi^0$ one.  Note that with the relation $\Amp(\eta_0\to 4\pi)
= \sqrt{2}\Amp(\eta_8\to 4\pi)$ and the standard mixing according to
Eq.~\eqref{eq:mixing},
we have $\Amp(\eta'\to 4\pi) = \Amp(\eta_8\to 4\pi)$. 
To
obtain branching ratios, we normalize the partial widths by the total width of the
$\eta'$ as quoted by the particle data group, $\Gamma_{\eta'} = (0.199\pm
0.009)\,{\rm MeV}$~\cite{PDG2010}. Note that by using the most precise single
measurement of this width alone, $\Gamma_{\eta'} = (0.226\pm 0.017 \pm 0.014)\,{\rm
MeV}$~\cite{Czerwinski}, our predictions for the branching fractions would be
reduced by more than 10\%. Given the observation of Eq.~\eqref{eq:VMvsKloop}, we
neglect the kaon-loop contributions altogether and evaluate the matrix elements
using Eq.~\eqref{eq:Arho1}. In order to account for trivial isospin-breaking
effects due to phase space corrections, we calculate the branching ratio for
$\eta'\to \pi^+\pi^0\pi^-\pi^0$ using an average pion mass $M_\pi=
(M_{\pi^+}+M_{\pi^0})/2$, while we employ the charged pion mass for the decay into
four charged pions. All results are first quoted as a function of the coupling
constants $c_1-c_2$ and $c_3$, before inserting two sets of values: (i) $c_1-c_2 =
c_3 =1$, and (ii) $c_1-c_2=1.21$, $c_3=0.93$~\cite{Benayoun}.
We refrain from employing the errors given in the fits in
Ref.~\cite{Benayoun}: the uncertainties in the HLS coupling constants are well
below what we estimate to be the overall uncertainty of our prediction.  The results are
\begin{align}
&\BR\big(\eta' \to 2(\pi^+\pi^-)\big) \nnnl
&= \Big[
  0.15 \,(c_1-c_2)^2
+ 0.47 \,(c_1-c_2)c_3
+ 0.37 \,c_3^2 \Big]\times 10^{-4} \nnnl
&= \big\{ 1.0 ,\, 1.1  \big\}\times 10^{-4} ~, \\
&\BR\big(\eta' \to \pi^+\pi^-2\pi^0\big) \nnnl
&= \Big[
  0.35  \,(c_1-c_2)^2
+ 1.09  \,(c_1-c_2)c_3
+ 0.87  \,c_3^2 \Big]\times 10^{-4} \nnnl
&= \big\{ 2.3 ,\, 2.5  \big\}\times 10^{-4} ~.
\end{align}
We therefore find that the uncertainties
due to the HLS coupling constants are small. We wish to point out that although
$\pi\pi$ P-wave dynamics are usually well approximated by the $\rho$ resonance, and
crossed-channel effects are expected to occur rather at the 10\% level (as inferred
from studies of decays such as $\omega \to 3\pi$, $\phi\to 3\pi$~\cite{Niecknig}),
the present study in some sense still amounts to a leading-order calculation:
SU(3)-breaking effects of the order of $F_\eta / F_\pi \approx
1.3$~\cite{GasserLeutwyler} may occur, and in the treatment of the $\eta'$
($\eta_0$), we have implicitly evoked the $1/N_c$ expansion.  We therefore deem a
generic uncertainty of 30\% realistic, and quote our predictions accordingly as
\begin{align}
\BR\big(\epTc \big) &= (1.0 \pm 0.3)\times 10^{-4} ~, \nnnl
\BR\big(\epTcn\big) &= (2.4 \pm 0.7) \times 10^{-4} ~.
\end{align}
These are to be compared to the current experimental upper limits~\cite{PDG2010,CLEO}
\begin{align}
\BR_{\rm exp}\big(\epTc \big) &< 2.4\times 10^{-4} ~, \nnnl
\BR_{\rm exp}\big(\epTcn\big) &< 2.6\times 10^{-3} ~,
\end{align}
hence signals of these decays ought to be within reach of modern high-statistics
experiments soon.

%--------------------------------------------------------------------------------
\section{\boldmath{$\eta,\,\eta'\to 4\pi^0$}}\label{sec:Neutral}
%--------------------------------------------------------------------------------

As we have mentioned in the Introduction, the P-wave mechanism described in the
previous section, proceeding essentially via two $\rho$ intermediate resonances,
cannot contribute to the $4\pi^0$ final states.  In fact, we can show that the
D-wave characteristic of $\eta,\,\eta' \to 4\pi^0$ suppresses these decays to
$\Order(p^{10})$ in chiral power counting, that is to the level of three loops in
the anomalous sector. This is, in particular, due to the flavor and isospin structure
of the anomaly, which does not contain five-meson vertices including $2\pi^0$ at
leading order ($\Order(p^4)$), and to the chiral structure of meson--meson
scattering amplitudes, which only allows for S and P~waves at tree level
($\Order(p^2)$). As a complete three-loop calculation would be a formidable task
and is certainly beyond the scope of our exploratory study, we instead consider the
decay mechanisms shown in Fig.~\ref{fig:4pi0loop}. As shown in
Appendix~\ref{app:tensor}, the contribution from two $f_2$ mesons is negligible in
comparison to the pion loop. We therefore focus on the pion loop as shown in the left
panel of Fig.~\ref{fig:4pi0loop}. It represents a decay mechanism that, we believe,
ought to capture at least the correct order of magnitude of the corresponding
partial width.

%--------------------------------------------------------------------------------
\subsection{Pion-loop contribution}
%--------------------------------------------------------------------------------

\begin{figure}
\centering \vskip 3mm
\includegraphics[width=0.5\linewidth]{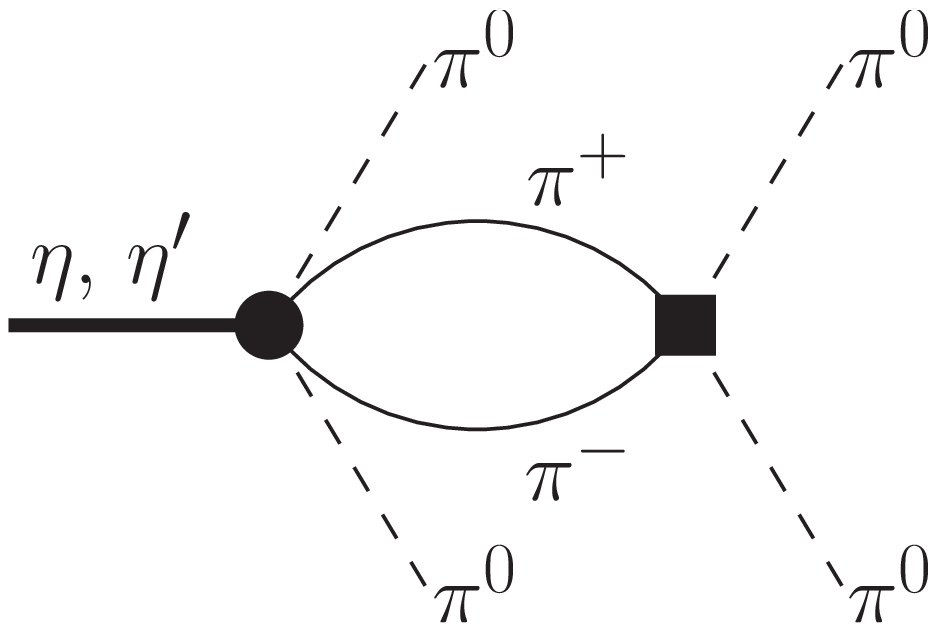}\hfill
\includegraphics[width=0.4\linewidth]{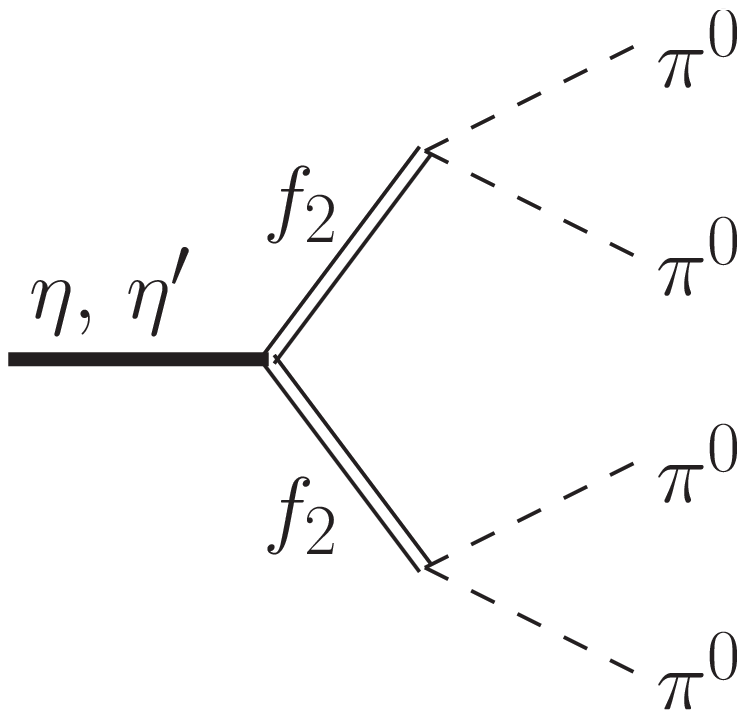}
\caption{Left: Pion-loop contribution to $\eta,\,\eta'\to 4\pi^0$.
The black circle denotes an effective local $\eta,\,\eta'\to \pi^+\pi^-2\pi^0$ coupling
at $\Order(p^6)$, the black square an effective local D-wave $\pi\pi$ scattering
vertex at $\Order(p^4)$.
Right: $\eta,\,\eta'\to 4\pi^0$ through two intermediate $f_2$ mesons.\label{fig:4pi0loop}}
\end{figure}
Our decay mechanism for $\eta,\,\eta' \to 4\pi^0$
is built on the observation that there is a specific diagrammatic contribution
that we can easily calculate, and that, in particular, comprises
the complete leading contribution to the \emph{imaginary part}
of the decay amplitude.
This is given by $\pi^+\pi^-$ intermediate states,
and hence harks back to the results of the previous section.
As argued above, it appears at chiral $\Order(p^{10})$:
$\eta_{0/8} \to  \pi^+\pi^- 2\pi^0$ as calculated
in Eq.~\eqref{eq:AmpOp6} to $\Order(p^6)$,
followed by rescattering $\pi^+\pi^- \to \pi^0\pi^0$, where
the S~wave does not contribute, and D and higher partial waves start to appear
at $\Order(p^4)$~\cite{GL-AnnPhys};  see Fig.~\ref{fig:4pi0loop} for illustration.
We calculate this first in the following approximation:
given the numerical dominance of the counterterm contribution in Eq.~\eqref{eq:AmpOp6},
the amplitudes $\F_{0/8}(s)$ are taken to be linear, $\F_{0/8}(s) \approx \F'_{0/8}(s)s$,
neglecting tiny curvature effects from the kaon loops;
and we approximate $\pi\pi$ rescattering by a phenomenological D~wave,
thus improving on the leading chiral representation, but neglecting G and higher partial waves.
We find
\begin{widetext}
\begin{align}
\Amp(\eta_8\to 4\pi^0) &= \frac{1}{\sqrt{2}} \Amp(\eta_0\to 4\pi^0)
= - \frac{N_c(c_1-c_2+c_3)}{8\pi}
\frac{\eps_{\mu\nu\alpha\beta}}{\sqrt{3}F_\pi^5}p_1^\mu p_2^\nu p_3^\alpha p_4^\beta
\Big\{
  \G(s_{12},s_{23},s_{14},s_{34};s_{13}) \nnnl &
+ \G(s_{12},s_{14},s_{23},s_{34};s_{24})
- \G(s_{13},s_{23},s_{14},s_{24};s_{12})
- \G(s_{13},s_{14},s_{23},s_{24};s_{34}) \nnnl &
- \G(s_{12},s_{24},s_{13},s_{34};s_{14})
- \G(s_{12},s_{13},s_{24},s_{34};s_{23})
\Big\} ~, \nnnl
\G(v,w,x,y;s) &= \frac{v-w-x+y}{M_\rho^2}\,\frac{16 (t_2^0(s)-t_2^2(s))}{3(s-4M_\pi^2)^2} \bigg\{
(s-4M_\pi^2)^2 \bar J_{\pi\pi}(s) \nnnl
&
- 2\big(s^2-10s M_\pi^2+30M_\pi^4\big) \Big(L + \frac{1}{16\pi^2}\log\frac{M_\pi}{\mu}\Big)
+ \frac{1}{16\pi^2} \bigg(
\frac{s^2}{15}-\frac{8}{3}s M_\pi^2 + 15M_\pi^4 \bigg) \bigg\} ~, \nnnl
\bar J_{\pi\pi}(s) &= \frac{1}{8\pi^2}\bigg\{1 - \frac{\sigma}{2}\Big(\log\frac{1+\sigma}{1-\sigma}-i\,\pi\Big)\bigg\} ~,
\quad \sigma = \sqrt{1-\frac{4M_\pi^2}{s}} ~, \quad
L = \frac{ \mu^{d-4}}{16\pi^2} \bigg\{ \frac{1}{d-4}+\frac{1}{2}(\gamma_E-1-\log 4\pi)\bigg\} ~.
\label{eq:A4pi0}
\end{align}
\end{widetext}
$t_2^I(s)$ is the partial wave of angular momentum $\ell=2$ for the appropriate
isospin quantum number $I$; the expression $16t_2^I(s)(s-4M_\pi^2)^{-2} = a_2^I +
\Order(s-4M_\pi^2)$, with the D-wave scattering length $a_2^I$, 
is therefore finite at threshold. Note furthermore that $t_2^I
= \Order(p^4)$ in chiral counting, such that the chiral order of
Eq.~\eqref{eq:A4pi0} is indeed $\Order(p^{10})$. 
$L$ contains the infinite part of the divergent loop diagram in the usual way, using dimensional regularization.
Of course, this individual loop
contribution is both divergent and scale-dependent: only the imaginary part is
complete (to this order) and in that sense well-defined and finite. We display the
full expression here as we will use the scale dependence as a rough independent
consistency check below.

Without the knowledge of counterterms of an order as high as $\Order(p^{10})$, one
cannot make a quantitative prediction using the loop amplitude derived in the
above.  Hence, we have to resort to a certain phenomenological representation.
The imaginary part of Eq.~\eqref{eq:A4pi0}, which is complete at
$\Order(p^{10})$ as mentioned, is used to establish a connection to a one-$f_2$
exchange in the $s$ channel. Note that the $f_2(1270)$ exchange dominates the
available $\pi\pi$ scattering phase shifts in the $I=0$, $\ell=2$ channel, see e.g.\ 
Ref.~\cite{Dobado:2001rv}.

We will proceed to estimate the full D-wave $\pi\pi$ rescattering contribution as follows. Neglecting again any
crossed-channel effects, rescattering of two pions can be summed by the
Omn\`es factor,
\be
\Omega_\ell^I(s) = \exp \bigg\{
\frac{s}{\pi} \int_{4M_\pi^2}^\infty \frac{\delta_\ell^I(z)dz}{z(z-s-i\eps)}
\bigg\}~, \label{eq:defOmnes}
\ee
where $\delta_\ell^I$ is the $\pi\pi$ scattering phase shifts in the channel with
isospin $I$ and angular momentum $\ell$. Near threshold, its imaginary part can be
approximated as
\be
\Im\Omega_\ell^I(s) \approx \delta_\ell^I(s) \big\{ 1+\Order(\sigma^2)\big\}
\approx \sigma \,t_\ell^I(s) \big\{ 1+\Order(\sigma^2)\big\} ~
\label{eq:ImOmnes_thr}
\ee
(neglecting the shift from unity in $\Omega(4M_\pi^2)$,
which is justified in the D~wave for our intended accuracy), while in the
approximation of a phase being dominated by a narrow resonance of mass $M$ and
width $\Gamma$, the Omn\`es factor is given by
\begin{align}
\Omega_\ell^I(s) &\approx \frac{M^2 \exp\left(i\delta_\ell^I(s)\right)}{\sqrt{(M^2-s)^2+M^2\Gamma^2(s)}} ~, \nnnl
\Gamma(s) &= \frac{M}{\sqrt{s}}\Big(\frac{s-4M_\pi^2}{M^2-4M_\pi^2}\Big)^{\ell+1/2} \Gamma ~.
\label{eq:Omnes_Res}
\end{align}
Despite $D$-wave scattering near threshold not being dominated by the
$f_2(1270)$,\footnote{It is dominated by the low-energy constant $\bar\ell_2$ from the $\Order(p^4)$
Lagrangian~\cite{GL-AnnPhys}, or by $t$-channel vector-meson exchange in the spirit of resonance
saturation~\cite{Ecker:1988te}.} we still use
Eq.~\eqref{eq:ImOmnes_thr} to invoke the $f_2$. This is because at somewhat higher
energies, the $I=0$ $\pi\pi$ D~wave dominates over the $I=2$ component and will be
well approximated by the $f_2(1270)$ resonance.  
One may wonder whether, in particular, for $\eTn$, which stays close to $\pi\pi$ threshold
throughout the allowed phase space, this approximation may not lead to sizeable errors.
We have checked for the numerical results for the branching fraction discussed below
that, employing the full Omn\`es function according to Eq.~\eqref{eq:defOmnes}
with the phase parameterization provided in Ref.~\cite{Madrid},
the branching ratio changes by about 10\%, well below the accuracy we can aim for here.
On the other hand, within the
$\eta'\to 4\pi^0$ decay, we stay sufficiently far below the resonance energy that
the phase of the D~wave can still be neglected. With the correspondence between
Eqs.~\eqref{eq:ImOmnes_thr} and~\eqref{eq:Omnes_Res}, we conclude that the
$f_2(1270)$ contribution to the amplitude can be estimated as
\begin{align}
\Amp_{f_2}(\eta_8\to 4\pi^0) &= \frac{1}{\sqrt{2}} \Amp_{f_2}(\eta_0\to 4\pi^0) \nnnl
&= - \frac{N_c(c_1-c_2+c_3)}{24\pi^2}
\frac{\eps_{\mu\nu\alpha\beta}}{\sqrt{3}F_\pi^5}p_1^\mu p_2^\nu p_3^\alpha p_4^\beta \nnnl
&\times \Big\{
  \G_{f_2}(s_{12},s_{23},s_{14},s_{34};s_{13},s_{24}) \nnnl
&\quad - \G_{f_2}(s_{13},s_{23},s_{14},s_{24};s_{12},s_{34})  \nnnl
&\quad - \G_{f_2}(s_{12},s_{24},s_{13},s_{34};s_{14},s_{23}) \Big\} ~, \nnnl
  \G_{f_2}(v,w,x,y;s,t) &= \frac{v-w-x+y}{M_\rho^2}  \nnnl &\times \bigg[
\frac{M_{f_2}^2}{M_{f_2}^2-s}+\frac{M_{f_2}^2}{M_{f_2}^2-t}\bigg] , \label{eq:Af2}
\end{align}
neglecting for simplicity the width of the $f_2$, which is justified in the kinematic
regime accessible in $\eta' \to 4\pi^0$.
Note that, due to the special symmetry of the amplitude, Eq.~\eqref{eq:Af2} can
be rewritten identically by employing a ``twice-subtracted'' version of the resonance
term, i.e.\ replacing $\G_{f_2} \to \G''_{f_2}$,
\begin{align}
\G''_{f_2}(v,w,x,y;s,t) &= \frac{v-w-x+y}{M_\rho^2 M_{f_2}^2} \nnnl &\times \bigg[
\frac{s^2}{M_{f_2}^2-s}+\frac{t^2}{M_{f_2}^2-t}\bigg] , \label{eq:Gf2''}
\end{align}
which makes the correct chiral dimension of the resonance contribution manifest.

As a rough final consistency check, we compare the order of magnitude of a chiral
counterterm induced by the $f_2$ exchange, see Eq.~\eqref{eq:Gf2''} in the
low-energy limit $s,\,t \ll M_{f_2}^2$ , with the scale running of such a counterterm as necessitated by
the $\log\mu$ dependence in Eq.~\eqref{eq:A4pi0}.  If we only retain the scattering
lengths in the D-wave partial waves, the relevant part to be compared to
Eq.~\eqref{eq:Gf2''} (that does not cancel in the full amplitude) is
\begin{align}
\mu \frac{d}{d\mu}&\big[\G(v,w,x,y;s)+\G(v,w,x,y;t)\big] \nnnl
&= \frac{v-w-x+y}{3M_\rho^2}\big(a_2^0-a_2^2\big) \frac{s^2 + t^2}{8\pi^2} ~.
\end{align}
Comparing the numerical prefactors, we find that the scale dependence is suppressed
versus the estimate for the finite counterterm by
\be
\frac{a_2^0-a_2^2}{16\pi} \times M_{f_2}^4 \approx 0.22 ~,
\ee
where we have used $a_2^0 = 1.75 \times 10^{-3}M_\pi^{-4}$, $a_2^2 = 0.17 \times 10^{-3}M_\pi^{-4}$~\cite{CGL}.
In other words, the scale dependence suggests the order of magnitude of our
counterterm estimate using $f_2$ saturation to be reasonable.

%--------------------------------------------------------------------------------
\subsection{Pion-loop contribution improved: including vector propagators}
%--------------------------------------------------------------------------------

\begin{figure}
\centering 
\includegraphics[width=0.5\linewidth]{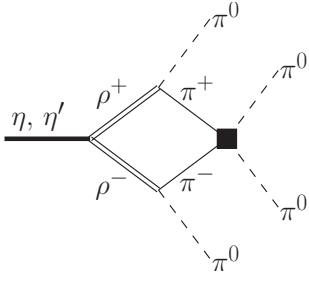}
\caption{Pion-loop contribution to $\eta,\,\eta'\to 4\pi^0$ via $\rho^\pm$ intermediate states;
see the vector-meson dominated amplitude discussed in Sec.~\ref{sec:Charged}.
The black square denotes an effective local D-wave $\pi\pi$ scattering
vertex at $\Order(p^4)$.\label{fig:4pi0viarho}}
\end{figure}

We have seen in Sec.~\ref{sec:Charged} on the P-wave dominated, (partially) charged
four-pion final states that the leading approximation in an expansion of the
$\rho$ meson propagators is not a sufficient description of these decays,
given the available phase space in $\eta'$ decays.
With the $\eta_{0/8} \to \pi^+\pi^-2\pi^0$ transitions entering
the decay mechanism for  $\eta_{0/8} \to 4\pi^0$ as described in the previous section,
this deficit would be fully inherited in our estimate of the all-neutral final states.
In fact, the imaginary part of the corresponding diagram including the full
$\rho$ propagators, see Fig.~\ref{fig:4pi0viarho}, can
even be calculated exactly, using Cutkosky rules; however, the resulting expressions are extremely
involved and not very illuminating.
It turns out, though, that the main effects of the not-so-large vector-meson mass can be
approximated by the following expression for the imaginary part:
\begin{widetext}
\begin{align}
\Im \Amp(\eta_8\to 4\pi^0) &= \frac{1}{\sqrt{2}} \Im \Amp(\eta_0\to 4\pi^0)
= - \frac{N_c}{8\pi}
\frac{\eps_{\mu\nu\alpha\beta}}{\sqrt{3}F_\pi^5}p_1^\mu p_2^\nu p_3^\alpha p_4^\beta
\Big\{ (c_1-c_2-c_3) \Big[
  \Im \G^\rho_1 (s_{12},s_{23},s_{14},s_{34};s_{13}) \nnnl & \quad
+ \Im \G^\rho_1(s_{12},s_{14},s_{23},s_{34};s_{24})
- \Im \G^\rho_1(s_{13},s_{23},s_{14},s_{24};s_{12})
- \Im \G^\rho_1(s_{13},s_{14},s_{23},s_{24};s_{34}) \nnnl & \quad
- \Im \G^\rho_1(s_{12},s_{24},s_{13},s_{34};s_{14})
- \Im \G^\rho_1(s_{12},s_{13},s_{24},s_{34};s_{23})\Big] \nnnl
& + 2 c_3 \Big[
  \Im \G^\rho_2(s_{12},s_{23},s_{14},s_{34};s_{13})
+ \Im \G^\rho_2(s_{12},s_{14},s_{23},s_{34};s_{24})
- \Im \G^\rho_2(s_{13},s_{23},s_{14},s_{24};s_{12}) \nnnl & \quad
- \Im \G^\rho_2(s_{13},s_{14},s_{23},s_{24};s_{34})
- \Im \G^\rho_2(s_{12},s_{24},s_{13},s_{34};s_{14})
- \Im \G^\rho_2(s_{12},s_{13},s_{24},s_{34};s_{23})
\Big] \Big\} ~, \nnnl
\Im \G^\rho_1(v,w,x,y;s) &=
\bigg[ \frac{M_\rho^2(v-w)}{\left(M_\rho^2-\frac{1}{2}(v+w)\right)^2}
-\frac{M_\rho^2(x-y)}{\left(M_\rho^2-\frac{1}{2}(x+y)\right)^2}\bigg]
\big(t_2^0(s)-t_2^2(s)\big) \frac{\sigma}{3\pi}
+ \Order\left(\sigma^7\right)~, \nnnl
\Im \G^\rho_2(v,w,x,y;s) &=
\frac{M_\rho^4\big(M_\rho^2(v-w-x+y) - vy + wx\big)}
     {\left(M_\rho^2-\frac{1}{2}(v+w)\right)^2\left(M_\rho^2-\frac{1}{2}(x+y)\right)^2} \,
\big(t_2^0(s)-t_2^2(s)\big)  \frac{\sigma}{3\pi}
+ \Order\left(\sigma^7\right)~.
\end{align}
We find, furthermore, that the neglected terms indicated as $\Order(\sigma^7)$ 
are also suppressed in inverse powers of $M_\rho$, starting at $\Order(M_\rho^{-6})$
compared to the leading terms of $\Order(M_\rho^{-2})$ in the above.
Numerically, the indicated higher-order corrections in $\sigma^2$ are found to be small,
less than about 10\% all over phase space.
However, the corrections by the remnants of the $\rho$ propagators are large
compared to the limit $M_\rho \to \infty$, given
the available phase space and the high power of these propagators in the denominator.
Using the same trick as in the previous section to transform the imaginary part
into an estimate for the whole (resonance-dominated) partial wave via the Omn\`es  function,
we arrive at
\begin{align}
\Amp(\eta_8\to 4\pi^0) &= \frac{1}{\sqrt{2}}  \Amp(\eta_0\to 4\pi^0)
= - \frac{N_c}{24\pi^2}
\frac{\eps_{\mu\nu\alpha\beta}}{\sqrt{3}F_\pi^5}p_1^\mu p_2^\nu p_3^\alpha p_4^\beta
\Big\{ (c_1-c_2-c_3) \Big[
  \G^\rho_{f_2,1} (s_{12},s_{23},s_{14},s_{34};s_{13}) \nnnl & \quad
+ \G^\rho_{f_2,1}(s_{12},s_{14},s_{23},s_{34};s_{24})
- \G^\rho_{f_2,1}(s_{13},s_{23},s_{14},s_{24};s_{12})
- \G^\rho_{f_2,1}(s_{13},s_{14},s_{23},s_{24};s_{34}) \nnnl & \quad
- \G^\rho_{f_2,1}(s_{12},s_{24},s_{13},s_{34};s_{14})
- \G^\rho_{f_2,1}(s_{12},s_{13},s_{24},s_{34};s_{23})
  \Big] \nnnl
&+ 2 c_3 \Big[
  \G^\rho_{f_2,2}(s_{12},s_{23},s_{14},s_{34};s_{13})
+ \G^\rho_{f_2,2}(s_{12},s_{14},s_{23},s_{34};s_{24})
- \G^\rho_{f_2,2}(s_{13},s_{23},s_{14},s_{24};s_{12}) \nnnl & \quad
- \G^\rho_{f_2,2}(s_{13},s_{14},s_{23},s_{24};s_{34})
- \G^\rho_{f_2,2}(s_{12},s_{24},s_{13},s_{34};s_{14})
- \G^\rho_{f_2,2}(s_{12},s_{13},s_{24},s_{34};s_{23})
\Big] \Big\} ~, \nnnl
\G^\rho_{f_2,1}(v,w,x,y;s) &=
\bigg[ \frac{M_\rho^2(v-w)}{\left(M_\rho^2-\frac{1}{2}(v+w)\right)^2}
-\frac{M_\rho^2(x-y)}{\left(M_\rho^2-\frac{1}{2}(x+y)\right)^2}\bigg]
 \frac{M_{f_2}^2}{M_{f_2}^2-s} ~, \nnnl
\G^\rho_{f_2,2}(v,w,x,y;s) &=
\frac{M_\rho^4\big(M_\rho^2(v-w-x+y) - vy + wx\big)}
     {\left(M_\rho^2-\frac{1}{2}(v+w)\right)^2\left(M_\rho^2-\frac{1}{2}(x+y)\right)^2} \,
 \frac{M_{f_2}^2}{M_{f_2}^2-s} ~. \label{eq:4pi0rhof2}
\end{align}
\end{widetext}
Note that this result is far from the one-$f_2$ dominance estimate, with a
$f_2$ coupling constant $\propto M_\rho^{-2}$ as the previous section suggested.
Expanding Eq.~\eqref{eq:4pi0rhof2} simultaneously around the limits $M_\rho \to
\infty$, $M_{f_2} \to \infty$, the leading term (corresponding to chiral dimension
$\Order(p^{10})$) is not dominated by terms of $\Order(M_\rho^{-2}M_{f_2}^{-4})$,
but also contains other terms of $\Order(M_\rho^{-4}M_{f_2}^{-2})$ and
$\Order(M_\rho^{-6})$. In other words, Eq.~\eqref{eq:Af2} is numerically no
reasonable approximation to Eq.~\eqref{eq:4pi0rhof2} even for the decay $\eta \to
4\pi^0$, with its tiny phase space available.

%--------------------------------------------------------------------------------
\subsection{Branching ratios}
\label{sec:numNeutral}
%--------------------------------------------------------------------------------

We calculate the partial width using Eq.~\eqref{eq:width} with the symmetry factor
$S=4!$. Note again that $\Amp(\eta'\to 4\pi^0) = \Amp(\eta_8\to 4\pi^0)$, assuming
standard mixing. We employ the amplitude as given in Eq.~\eqref{eq:4pi0rhof2} as
our ``best guess'' for an estimate of the branching fraction. With the same
numerical input as in Sec.~\ref{sec:numCharged} (except using the \emph{neutral}
pion mass everywhere), we find
\begin{align}
&\BR\big(\epTn \big) \nnnl &= \big[
  0.4 \,(c_1-c_2)^2
+ 1.6 \,(c_1-c_2)c_3
+ 1.7 \, c_3^2 \big]\times 10^{-8} \nnnl &
= \big\{ 3.7 ,\, 3.9 \big\}\times 10^{-8} ~,\label{eq:etaprime4pi0width}
\end{align}
for the two sets of coupling constants $c_i$.
Note that the use of the amplitude~\eqref{eq:Af2} leads to a branching fraction
of the order of $4\times 10^{-11}$, i.e.\ almost 3 orders of magnitude smaller.

We can trivially also calculate the branching fraction for $\eta \to 4\pi^0$, the
only $\eta \to 4\pi$ decays that is kinematically allowed. We again employ the
amplitude~\eqref{eq:4pi0rhof2}, and note that mixing according to
Eq.~\eqref{eq:mixing} suggests $\Amp(\eta\to 4\pi^0) = \sqrt{2}\Amp(\eta_8\to
4\pi^0)$. Normalized to the total width of the $\eta$, $\Gamma_\eta = (1.30 \pm
0.07)\,{\rm keV}$~\cite{PDG2010}, we find
\begin{align}
&\BR\big(\eTn \big) \nnnl &= \big[
  0.4 \,(c_1-c_2)^2
+ 1.1 \,(c_1-c_2)c_3
+ 1.0 \, c_3^2 \big]\times 10^{-30} \nnnl
&= \big\{ 2.4 ,\, 2.6 \big\}\times 10^{-30} ~, \label{eq:eta4pi0width}
\end{align}
in other words, the D-wave characteristic of the decay combined with tiny phase space
leads to an enormous suppression of the $CP$-allowed $\eta\to 4\pi^0$ decay.
We again compare these estimates to the available experimental upper limits~\cite{Alde,Prakhov},
\begin{align}
\BR_{\rm exp}\big(\epTn\big) &< 5\times 10^{-4} ~, \nnnl
\BR_{\rm exp}\big(\eTn \big) &< 6.9\times 10^{-7} ~;
\end{align}
further improvements of these experimental upper limits are planned
(see e.g.\ Ref.~\cite{Bednarski} for $\eTn$).
In this case, our predictions are smaller than those by several orders of magnitude.

The uncertainties of Eqs.~\eqref{eq:etaprime4pi0width} and \eqref{eq:eta4pi0width}
are hard to assess.  The generic SU(3) and $1/N_c$ error of about 30\% assumed in
Sec.~\ref{sec:numCharged} is probably too small, as here, we do not even have a
complete leading-order calculation at our disposal. We therefore rather assume
these numbers to be the correct orders of magnitude, without quantifying the
uncertainty of the prediction any further.

%--------------------------------------------------------------------------------
\section{\boldmath{$CP$}-violating \boldmath{$\eta,\,\eta'\to 4\pi^0$} decays}\label{sec:CP}
%--------------------------------------------------------------------------------

Given the smallness of the branching fractions predicted for $\epTn$, $\eTn$ via a
D-wave dominated, $CP$-conserving decay mechanism in the previous section, it is
desirable to compare these numbers with possible $CP$-violating contributions that
may, on the other hand, avoid the huge angular-momentum suppression. One  such
$CP$-violating mechanism that is expected to affect strong-interaction processes is
induced by the so-called $\theta$-term, an additional term in the QCD Lagrangian
necessitated for the solution of the U(1)$_A$ problem. The $\theta$-term
violates $P$ and $CP$ symmetry and may induce observable symmetry-violating effects, in
particular, in flavor-conserving processes. Its effective-Lagrangian treatment
includes a term that can be rewritten as (see Ref.~\cite{PichRafael} and references therein)
\begin{align}
\Lagr_{\theta} &= i\,\bar\theta_0\,\frac{F_\pi^2 M_{\eta_0}^2 }{12}  \bigg\{
\langle U-U^\dagger\rangle - \log\Big(\frac{\det U}{\det U^\dagger}\Big) \bigg\} ~, \nnnl
U &= u^2 = \exp\Big( \frac{i \varphi}{F_\pi} \Big) ~,
\end{align}
which, in addition to the well-known $\eta \to 2\pi$ amplitude~\cite{CVVW,PichRafael},
also induces a $CP$-violating $\eta \to 4\pi$ amplitude,
\begin{align}
& \Amp_{CP}(\eta_8\to 4\pi^0) = \frac{1}{\sqrt{2}}\Amp_{CP}(\eta_0\to 4\pi^0)
\nnnl &= \Amp_{CP}(\eta'\to 4\pi^0) = \frac{1}{\sqrt{2}}\Amp_{CP}(\eta\to 4\pi^0)
= -\frac{M_{\eta_0}^2 \bar\theta_0}{3\sqrt{3}F_\pi^3} ~.
\end{align}
We will use $M_{\eta_0} \approx \mep$ for numerical evaluation.
The fact that this amplitude is a constant makes the phase space integration
almost trivial, with the results for the branching fractions
\begin{align}
\BR(\eta \stackrel{CPV}{\longrightarrow} 4\pi^0) &= 5 \times 10^{-5} \times \bar\theta_0^2 ~, \nnnl
\BR(\eta' \stackrel{CPV}{\longrightarrow} 4\pi^0) &= 9 \times 10^{-2} \times \bar\theta_0^2 ~.
\label{eq:CPBR}
\end{align}
We remark that we do not consider the branching ratio estimate for $\eta'\to 4\pi^0$
in Eq.~\eqref{eq:CPBR} reliable in any sense: given the available phase space and
the possibility of strong S-wave $\pi\pi$ final-state interactions, it could easily be enhanced
by an order of magnitude.
Were $\bar\theta_0$ a quantity of natural size, Eq.~\eqref{eq:CPBR}
would demonstrate the enhancement of the
$CP$-violating S-wave mechanism compared to the $CP$-conserving D-wave one,
see Eqs.~\eqref{eq:etaprime4pi0width} and \eqref{eq:eta4pi0width}.
With current limits on the QCD vacuum angle derived
from neutron electric dipole moment measurements, $\bar\theta_0 \lesssim 10^{-11}$~\cite{Ottnad},
these branching fractions are  already bound beyond anything measurable;
however, we note that for $\eta \to 4\pi^0$, the suppression of the $CP$-conserving D-wave
mechanism, see Eq.~\eqref{eq:eta4pi0width}, is so strong that it is even smaller
than the $CP$-violating (S-wave) one in Eq.~\eqref{eq:CPBR} if  the current bounds are inserted for $\bar\theta_0$.

%--------------------------------------------------------------------------------
\section{Summary and conclusions}
%--------------------------------------------------------------------------------

In this article, we have calculated the branching fractions of the $\eta$
and $\eta'$ decays into four pions. 
These processes of odd intrinsic parity are anomalous, and---as long
as $CP$ symmetry is assumed to be conserved---forbid the pions
to be in relative S-waves.
We organize the amplitudes according to chiral power-counting rules,
and find the leading contributions to the $\eta'$ decay amplitudes with 
charged pions in the final state at $\Order(p^6)$.
Utilizing the framework of hidden local symmetry for vector mesons, we
assume that vector-meson exchange saturates the $\Order(p^6)$ low-energy
constants, and find that the (P-wave) decay amplitude is entirely governed 
by $\rho$ intermediate states.
The dominant contribution is hence given by the triangle anomaly
via $\eta' \to \rho\rho$ (with numerically subleading box terms),
not by the pentagon anomaly.
In this way, the branching fractions for $\epTc$ and $\epTcn$ are
predicted to be 
\begin{align}
\BR\big(\epTc \big) &= (1.0\pm0.3)\times10^{-4} ~, \nnnl
\BR\big(\epTcn \big) &= (2.4\pm0.7)\times10^{-4} ~,
\end{align}
respectively. The former is only a factor of 2 smaller than the current
experimental upper limit, so should be testable in the near future with the modern
high-statistics facilities. 

Predictions for the decays into four neutral pions are much more difficult, 
as Bose symmetry requires them to emerge in relative D-waves (assuming $CP$ conservation),
suppressing the amplitudes to $\Order(p^{10})$ in chiral power counting.
We here do not even obtain the full leading-order amplitudes, as these
would require a three-loop calculation.
We estimate the decay via a charged-pion-loop contribution with D-wave
pion--pion charge-exchange rescattering;
an alternative mechanism through two $f_2$ mesons is found to be 
completely negligible in comparison, based on an 
estimate of the tensor--tensor--pseudoscalar coupling constant in the framework of
QCD sum rules.
Because of these phenomenological approximations, the $CP$-conserving branching ratios thus obtained,
\begin{align}
\BR\big(\epTn \big)  &\sim 4 \times 10^{-8} ~, \nnnl
\BR\big(\eTn \big)  &\sim 3 \times 10^{-30} ~,
\end{align}
should only be taken as order-of-magnitude estimates. 
It thus turns out that the $CP$-conserving decay width of $\eTn$ is so small that any signal to
be observed would indicate $CP$-violating physics.
For the latter, we calculate one specific example using the QCD $\theta$-term.

\begin{acknowledgments}
We would like to thank Andrzej Kup\'s\'c for initiating this project and for discussions,
and Maurice Benayoun for useful communications concerning Ref.~\cite{Benayoun}.
Partial financial support by  the Helmholtz Association through funds provided to
the Virtual Institute ``Spin and strong QCD'' (VH-VI-231), 
by the DFG (SFB/TR 16, ``Subnuclear Structure of Matter''),
and by the project ``Study of Strongly Interacting Matter'' 
(HadronPhysics2, Grant No.\ 227431) under the Seventh Framework Program of the EU
is gratefully acknowledged.
\end{acknowledgments}

\appendix

%--------------------------------------------------------------------------------
\section{Four-body phase space integration}
\label{app:ps}
%--------------------------------------------------------------------------------

The $n$-body phase space is defined as
\begin{align}
&d\Phi_n(P;p_1,\dots,p_n) \nnnl &\equiv (2\pi)^4 \delta^4\Big(P-\sum_{i=1}^{n}p_i\Big)
\prod_{i=1}^{n} \frac{d^3p_i}{(2\pi)^3 2 p_i^0} ~.
\end{align}
Using the recursive relation~\cite{PDG2010}
\begin{align}
d\Phi_n(P&;p_1,\dots,p_n) = d\Phi_j(q;p_1,\dots,p_j) \nnnl & \times
d\Phi_{n-j+1}(P;q,p_{j+1},\dots,p_n) \frac{dq^2}{2\pi} ~, 
\end{align}
we have
\begin{align}
\label{eq:ps4}
&d\Phi_4(P;p_1,\dots,p_4) \nnnl &= d\Phi_2(q;p_1,p_2)
d\Phi_2(k;p_3,p_4)
d\Phi_2(P;q,k) \frac{dq^2}{2\pi} \frac{dk^2}{2\pi} \nnnl
&= \frac1{(8\pi^2)^4 M} \int_{m_1+m_2}^{M-m_3-m_4} \!\!\! d\sqrt{s_{12}}
\int_{m_3+m_4}^{M-\sqrt{s_{12}}}\!\!\! d\sqrt{s_{34}} \nnnl
& \qquad \times \int d\Omega_{1}^*d\Omega_{3}'d\Omega
|{\bm p}_1^*| |{\bm p}_3'| |{\bm q}| ~,
\end{align}
where $m_i$, $i=1\ldots 4$ are the masses associated with the final-state particles
of momentum $p_i$, $M$ is the mass of the decaying particle,
$s_{12}=q^2$, $s_{34}=k^2$.  
$d\Omega_{1}^*=d\varphi_1^* d\cos\theta_1^*$ is the solid angle of particle 1 in the center-of-mass frame
(cmf) of particles 1 and 2, $d\Omega_{3}'$ is the solid angle of particle 3 in the cmf of
3 and 4, and $d\Omega$ is the solid angle of the 1,\,2 system in the rest frame of
the decaying particle. The three-momenta are given by
\begin{align}
|{\bm p}_1^*| &= \frac{\lambda^{1/2}(s_{12},m_1^2,m_2^2)}{2\sqrt{s_{12}}} ~, \quad
|{\bm p}_3'| = \frac{\lambda^{1/2}(s_{34},m_3^2,m_4^2)}{2\sqrt{s_{34}}} ~, \nnnl
|{\bm q}| &= \frac{\lambda^{1/2}(M^2,s_{12},s_{34})}{2M} ~,
\end{align}
with the usual K\"all\'en function $\lambda(x,y,z)\equiv x^2+y^2+z^2-2(x y+x z+y z)$.

Denoting quantities in the cmf of 1 and 2 (3 and 4) by $^*(')$, one can relate
them with those in the rest frame of the decay particle by Lorentz
transformation. Explicitly,
\begin{align}
p_1^\mu &= \{\gamma_{12}(p_1^{0*}+{\bm \beta}_{12}\cdot{\bm p}_1^*),
\gamma_{12} ({\bm \beta}_{12}p_1^{0*} + {\bm p}_{1\parallel}^*), {\bm p}_{1\perp}^* \} ~,\nnnl
p_2^\mu  &= \{\gamma_{12}(p_2^{0*}-{\bm \beta}_{12}\cdot{\bm p}_1^*),
\gamma_{12} ({\bm \beta}_{12}p_2^{0*} - {\bm p}_{1\parallel}^*), -{\bm p}_{1\perp}^* \} ~, \nnnl
p_3^\mu &= \{\gamma_{34}(p_3^{0\prime}+{\bm \beta}_{34}\cdot{\bm p}_3'),
\gamma_{34} ({\bm \beta}_{34}p_3^{0\prime} + {\bm p}_{3\parallel}'), {\bm p}_{3\perp}' \} ~, \nnnl
p_4^\mu &= \{\gamma_{34}(p_4^{0\prime}-{\bm \beta}_{34}\cdot{\bm p}_3'),
\gamma_{34} ({\bm \beta}_{34}p_4^{0\prime} - {\bm p}_{3\parallel}'), -{\bm p}_{3\perp}' \} ~,
\end{align}
where ${\bm \beta}_{12} = {\bm q}/q^0$ $({\bm \beta}_{34} = {\bm k}/k^0)$ is the
velocity of the 1,\,2 (3,\,4) system in the rest frame of the decay particle, and
$\gamma_{12(34)}=(1-{\bm \beta}_{12(34)}^2)^{-1/2}$. Moreover,
${\bm p}_{1\parallel(\perp)}^*$ are the components of ${\bm p}_1^*$ parallel
(perpendicular) to ${\bm q}$, and ${\bm p}_{3\parallel(\perp)}'$ are the
components of ${\bm p}_3'$ parallel (perpendicular) to ${\bm k}$.
\begin{figure}
\centering
\includegraphics[width=0.9\linewidth]{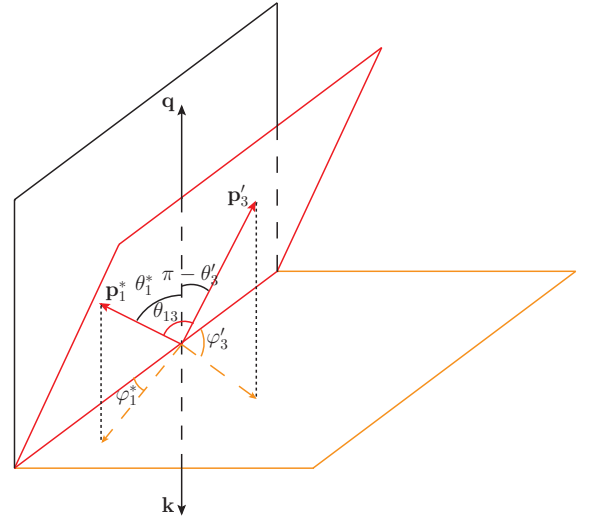}
\caption{The solid angles of particle 1 (3) in the cmf of 1 and 2 (3 and 4).
$\theta_{13}$ is the angle between the momentum of particle 1 in the cmf of 1 and 2 and
the momentum of particle 3 in the cmf of 3 and 4.\label{fig:angles}}
\end{figure}
One can define $\theta_1^*$ as the angle between the directions of ${\bm q}$ and
${\bm p}_1^*$, and $\theta_3'$ as the one between ${\bm k}=-{\bm q}$ and ${\bm
p}_3'$. The angle between ${\bm p}_1^*$ and ${\bm p}_3'$, $\theta_{13}$, is
related to the solid angles $\Omega_1^*$ and $\Omega_3'$ by
\be
\cos\theta_{13} = - \cos\theta_1^*\cos\theta_3' -
\sin\theta_1^*\sin\theta_3'\cos(\varphi_3'+\varphi_1^*)~.
\ee
The angles are shown for illustration in Fig.~\ref{fig:angles}.
It is obvious that the integration $d\Omega$ as well as the one over either $\varphi_1^*$
or $\varphi_3'$ are trivial, such that Eq.~\eqref{eq:ps4} simplifies to
\begin{align}
&d\Phi_4(P;p_1,\dots,p_4) \nnnl
&= \frac1{(8\pi^2)^3 M} \int_{m_1+m_2}^{M-m_3-m_4} \!\!\! d\sqrt{s_{12}}
\int_{m_3+m_4}^{M-\sqrt{s_{12}}}\!\!\! d\sqrt{s_{34}} \nnnl
& \qquad \times\int d\cos\theta_1^*d\cos\theta_3'd\varphi_3'
|{\bm p}_1^*| |{\bm p}_3'| |{\bm q}| ~.
\end{align}

%--------------------------------------------------------------------------------
\section{Tensor-meson contributions to \boldmath{$\eta,\,\eta'\to 4\pi^0$}}
\label{app:tensor}
%--------------------------------------------------------------------------------

%--------------------------------------------------------------------------------
\subsection{Amplitude, decay width}
%--------------------------------------------------------------------------------

In this Appendix, we discuss an alternative, resonance-driven decay mechanism for
the decays $\eta,\,\eta'\to 4\pi^0$, namely, via two $f_2(1270)$ tensor mesons, see
the right panel of Fig.~\ref{fig:4pi0loop}. The Lagrangian for the
tensor--tensor--pseudoscalar interaction reads 
\be 
\Lagr_{TTP} =
\frac{g_{TTP}}{\sqrt{2}} 
\eps_{\mu\nu\rho\sigma}g_{\alpha\beta}
\langle\partial^\mu T^{\nu\alpha} T^{\rho\beta} \partial^\sigma \varphi \rangle ,
\ee
where the tensor nonet is given by
\begin{align}
T_{\mu\nu} &= \left(
\begin{array}{ccc}
\frac{a_{2\mu\nu}^0}{\sqrt{2}} + \frac{f_{2\mu\nu}}{\sqrt{2}} & a_{2\mu\nu}^+ & K_{2\mu\nu}^+\\[2mm]
a_{2\mu\nu}^- & -\frac{a_{2\mu\nu}^0}{\sqrt{2}} + \frac{f_{2\mu\nu}}{\sqrt{2}} & K_{2\mu\nu}^0\\[2mm]
K_{2\mu\nu}^- & \bar{K}_{2\mu\nu}^0 & f_{2\mu\nu}'
\end{array}
\right) .
\end{align}
The coupling constant $g_{TTP}$ is not easily determined phenomenologically;
we will first write the resulting decay width as a function of $g_{TTP}$, and then
proceed to estimate it in Appendix~\ref{app:qcdsr} using QCD sum rules.

The decay
of the $f_2$ into two pseudoscalars is described by the
Lagrangian~\cite{Donoghue:1988ed,Dobado:2001rv}
\be
\Lagr_{f_2} = g_T f_{2\mu\nu}  \langle u^\mu u^\nu \rangle
= \frac{g_T}{F_\pi^2} f_{2\mu\nu}  \langle \partial^\mu \varphi \partial^\nu \varphi\rangle
+ \ldots ~.
\ee
Using the polarization sum for a tensor meson~\cite{Novozhilov}
\be
\label{eq:polsum} \sum_{\lambda} \phi_{\mu\nu}^{(\lambda)}(p)
\phi_{\rho\sigma}^{(\lambda)}(p)^\dag = \frac12 (X_{\mu\rho}X_{\nu\sigma} +
X_{\mu\sigma}X_{\nu\rho}) - \frac1{3} X_{\mu\nu}X_{\rho\sigma} ~,
\ee
with
$X_{\mu\nu}\equiv g_{\mu\nu}-p_\mu p_\nu / M_{f_2}^2$, it is straightforward to
derive the $f_2\to\pi\pi$ decay width as~\cite{Dobado:2001rv}
\be
\Gamma (f_2\to\pi\pi)=
\frac{g_T^2}{80\pi M_{f_2}^2F_\pi^4} \left(M_{f_2}^2-4M_\pi^2\right)^{5/2}.
\ee
Inserting $M_{f_2}=1275.1$\,MeV, $\Gamma_{f_2} = 185.1$\,MeV, and
$\BR(f_2\to\pi\pi)=84.8\%$~\cite{PDG2010} (neglecting the corresponding
uncertainties), the coupling constant can be obtained as $g_T = 39.4$\,MeV.

Applying the polarization sum of a tensor field as given in Eq.~\eqref{eq:polsum}, the
decay amplitude for the process $\eta'\to f_2f_2\to4\pi^0$ takes the simple form
\begin{align}
\Amp_{f_2f_2} &= - \sqrt{\frac{2}{3}} \frac{2g_{TTP}g_T^2}{F_\pi^4}
\eps_{\mu\nu\alpha\beta} p_1^\mu p_2^\nu p_3^\alpha p_4^\beta \nnnl
&\times
\big[\mathcal{H}(s_{12},s_{23},s_{14},s_{34};s_{13},s_{24}) \nnnl
&\quad -\mathcal{H}(s_{13},s_{23},s_{14},s_{24};s_{12},s_{34})\nnnl
&\quad -\mathcal{H}(s_{12},s_{24},s_{13},s_{34};s_{14},s_{23}) \big] ~. \label{eq:AgTTP}
\end{align}
The three terms in the square brackets are due
to interchange of identical pions in the final state, and $\mathcal{H}(v,w,x,y;s,t)$
is given by
\be
\mathcal{H}(v,w,x,y;s,t) = \frac{v-w-x+y} {(M_{f_2}^2-s)
(M_{f_2}^2-t)} ~,
\ee
where we have again neglected the $f_2$ width in the propagators.

We calculate the partial width as in Sec.~\ref{sec:numNeutral},
and find as the result for the decay mechanism through two virtual $f_2$ states
\be
\Gamma(\eta'\to f_2f_2\to4\pi^0) \approx 1\times10^{-16} \frac{g_{TTP}^2}{{\mbox{\small
GeV}}^{-2}}\,{\rm MeV}. \vspace{3mm}
\ee
The value of $g_{TTP}$ is estimated using the method of QCD sum rules in
Appendix~\ref{app:qcdsr} to be about 9\,GeV$^{-1}$.  Thus, the branching fraction
is
\be
\BR(\eta'\to f_2f_2\to4\pi^0) \approx 4\times10^{-14} ~.
\ee
It is orders of magnitude smaller than the value in
Eq.~\eqref{eq:etaprime4pi0width}, and hence can be safely neglected.

%--------------------------------------------------------------------------------
\subsection{Estimate of \boldmath{$g_{TTP}$} via QCD sum rules}
\label{app:qcdsr}
%--------------------------------------------------------------------------------

In this Appendix, we estimate the unknown coupling constant $g_{TTP}$
using QCD sum rules~\cite{Shifman:1978bx,Reinders:1984sr}. We choose to estimate it
from the $a_2f_2\pi$ coupling. Since we are not aiming at a precise calculation,
complications due to mixing with gluon operators and anomalous dimensions will be
neglected. The interpolating fields for the $a_2^+$~\cite{Aliev:1981ju} and $\pi^+$
are
\begin{align}
j_{\mu\nu}^{(a_2^+)}(x) &= \frac{i}{2} \bar d(x)
(\gamma_\mu\overleftrightarrow{D}_{\nu} + \gamma_\nu\overleftrightarrow{D}_{\mu})
u(x) ~, \nnnl
j_5^{(\pi^+)}(x) &= i m_q\bar d(x) \gamma_5 u(x) ~,
\end{align}
where
$\overleftrightarrow{D}_{\mu}\equiv(\overrightarrow{D}_\mu-\overleftarrow{D}_\mu)/2$,
with $D_\mu$ the standard covariant derivative,
and $m_q$ is the light quark mass. The flavor wave function for the $f_2$ is $(\bar
u u+\bar d d)/\sqrt{2}$, and the corresponding interpolating field follows from the
above equation. We will study the three-point correlation function
\begin{widetext}
\begin{align}
\Pi_{\mu\nu\alpha\beta}(p',q) &= \int d^4x d^4y e^{i(p'x+qy)}
\Big\langle0\Big| T\Big\{ j_{\mu\nu}^{(f_2)}(x) j_5^{(\pi^+)}(y)
\Big[j_{\alpha\beta}^{(a_2^+)}(z)\Big]^\dag \Big\} \Big|0\Big\rangle_{z\to0} \nnnl
 &\equiv \Pi(p',q) p^{\prime\rho}q^\sigma \left(
 g_{\alpha\mu}\eps_{\beta\nu\rho\sigma} + g_{\alpha\nu}\eps_{\beta\mu\rho\sigma}
 + g_{\beta\mu}\eps_{\alpha\nu\rho\sigma}  + g_{\beta\nu}\eps_{\alpha\mu\rho\sigma}
 \right) + \ldots ~.
\end{align}
\end{widetext}
In the last step, only one Lorentz structure is kept, which we will use for the estimate.

The operator product expansion for the correlation function can be calculated in
the deep Euclidean region. Keeping operators of the lowest dimension only, we find
\be 
\Pi(p',q) = - \frac{m_q\langle\bar qq\rangle}{16\sqrt{2}} \frac1{q^2}
\left[\log(-p^2)+\log(-p^{\prime2})\right] ~,
\ee 
with $p=q+p'$, where we have neglected
all polynomial terms since they will not contribute after the Borel transform. The
correlation function can also be expressed in terms of hadronic quantities, which
reads 
\be 
\Pi(p',q) = - \frac1{4} F_\pi M_\pi^2\frac{g_{TTP} f_T^2
M_T^6}{(q^2-M_\pi^2)(p^2-M_T^2)(p^{\prime2}-M_T^2)} ~, 
\ee 
where SU(3) symmetry is
assumed for the $a_2$ and $f_2$ by requiring their decay constants and masses to be
the same. The hadronic quantities are defined as
\be
\langle0|j_5^{(\pi^+)}|\pi^+\rangle = \frac1{\sqrt{2}}F_\pi M_\pi^2 ~,\quad
\langle0|j_{\mu\nu}^{(f_2)}|f_2\rangle = f_T M_T^3 \phi_{\mu\nu} ~,
\ee 
with $\phi_{\mu\nu}$ as defined in Eq.~\eqref{eq:polsum}.
\begin{figure}
\centering
\includegraphics[width=0.9\linewidth]{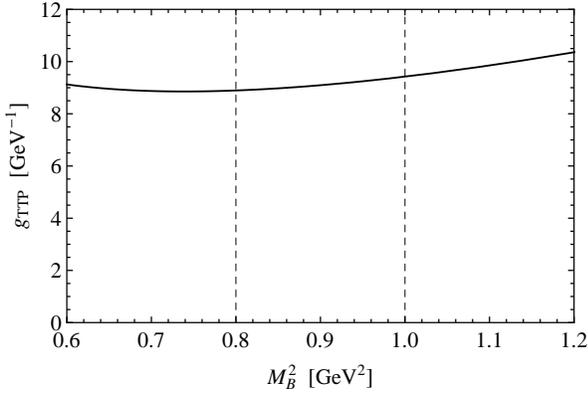}
\caption{QCD sum rule estimate of $g_{TTP}$. The interval
$M_B^2\in[0.8,1.0]$\,GeV$^2$, marked by the vertical dashed lines,
is optimal to suppress both higher power corrections and
continuum contribution in Ref.~\cite{Aliev:1981ju}. \label{fig:gttp}}
\end{figure}
Note that $f_T$ defined in this way is dimensionless. Following
Ref.~\cite{Reinders:1982hd}, we take $p^2=p^{\prime2}$, and perform the Borel
transform only once. We obtain the sum rule 
\be 
g_{TTP} = \frac{F_\pi M_B^4}{4\sqrt{2} f_T^2M_T^6} e^{M_T^2/M_B^2} ~, 
\ee 
where $M_B$ is the so-called Borel mass, and
$m_q\langle\bar q q\rangle=-M_\pi^2F_\pi^2/2$ has been used. In addition, $f_T$ was
already calculated in QCD sum rules~\cite{Aliev:1981ju}. Because of a cancellation
between the gluon condensate and the four-quark condensate, the sum rule for $f_T$
is dominated by a perturbative contribution, which reads~\cite{Aliev:1981ju} 
\be
M_T^6 f_T^2 e^{-M_T^2/M_B^2} \approx \frac3{160\pi^2} \int_0^{s_0} s^2 e^{-s/M_B^2} ds, 
\ee 
where $s_0$ is a threshold parameter introduced to mimic the spectral
function in the region $q^2>s_0$ by the one calculated using perturbative QCD.
Finally, we obtain
\be 
g_{TTP} \approx \frac{20\sqrt{2}\pi^2}{3}F_\pi M_B^4 \left( \int_0^{s_0}
s^2 e^{-s/M_B^2} ds \right)^{-1}. \label{eq:gttp}
\ee 
This estimate is plotted in
Fig.~\ref{fig:gttp} as a function of $M_B^2$. Taking the same interval of
$M_B^2\in[0.8,1.0]$\,GeV$^2$ and $s_0=2.5$\,GeV$^2$ as in
Ref.~\cite{Aliev:1981ju}, the coupling constant is estimated as 
\be 
g_{TTP} \approx 9\,{\rm GeV}^{-1}~. \label{eq:Ngttp} 
\ee

%--------------------------------------------------------------------------------

%--------------------------------------------------------------------------------

\end{document}